\documentclass{aastex61}

\received{...}
\revised{...}
\accepted{...}
\submitjournal{ApJ}

\shorttitle{Influence of ice on a water-forming reaction}
\shortauthors{Lamberts and K\"astner}

\usepackage{color}
\usepackage[english]{babel}
\usepackage[version=3]{mhchem}
\usepackage{rotating}

\begin{document}

\title{Influence of surface and bulk water ice on the reactivity of a water-forming reaction }
\correspondingauthor{Thanja Lamberts}
\email{lamberts@theochem.uni-stuttgart.de}

\author[0000-0001-6705-2022]{Thanja Lamberts}
\affil{Institute for Theoretical Chemistry \\ University Stuttgart \\ Pfaffenwaldring 55 \\ 70569 Stuttgart \\ Germany}

\author[0000-0001-6178-7669]{Johannes K\"astner}
\affil{Institute for Theoretical Chemistry \\ University Stuttgart \\ Pfaffenwaldring 55 \\ 70569 Stuttgart \\ Germany}

\keywords{ Astrochemistry - Molecular processes - ISM: molecules  }

\begin{abstract}
{On the surface of icy dust grains in the dense regions of the interstellar medium a rich chemistry can take place. Due to the low temperature, reactions that proceed via a barrier can only take place through tunneling. The reaction \ce{H + H2O2 -> H2O + OH} is such a case with a gas-phase barrier of $\sim$26.5 kJ/mol. Still the reaction is known to be involved in water formation on interstellar grains. Here, we investigate the influence of a water ice surface and of bulk ice on the reaction rate constant. Rate  constants are  calculated  using  instanton  theory down to 74~K. The ice is taken into account via multiscale modeling, describing the reactants and the direct surrounding at the quantum mechanical level with density functional theory (DFT), while the rest of the ice is modeled on the molecular mechanical level with a force field. We find that \ce{H2O2} binding energies cannot be captured by a single value, but rather depend on the number of hydrogen bonds with surface molecules. In highly amorphous surroundings the binding site can block the routes of attack and impede the reaction. Furthermore, the activation energies do not correlate with the binding energies of the same sites. The unimolecular rate constants related to the Langmuir-Hinshelwood mechanism increase as the activation energy decreases. Thus, we provide a lower limit for the rate constant and argue that rate constants can have values up to two order of magnitude larger than this limit. }
\end{abstract}

\section{Introduction}

In the dark molecular clouds of the interstellar medium (ISM) exothermic chemical reactions can take place despite the low temperature and densities present. Dust grains covered with water ice namely provide a meeting place for reactants as well as a third body to take up excess energy of the reaction.

One of the pathways leading to water formation on dust grains in dense regions of the ISM is 
\begin{equation}
 \ce{H + H2O2} \xrightarrow{k} \ce{H2O + OH} \;\; \tag{R1}\label{R1} .
\end{equation}
This reaction can take place after \ce{H2O2} has been formed via the subsequent addition of 2 hydrogen atoms to an oxygen molecule or via the recombination of two OH radicals. Reaction~\ref{R1} is known to have a high gas-phase barrier, see \citet{Baulch:2005} and references therein. Experiments on the surface at 15~K, however, have shown the reaction to take place nonetheless \citep{Oba:2014}, indicating the importance of tunneling. 

Besides the fact that hydrogen peroxide can lead to the formation of water, it has also been given recent attention as a result of the gas-phase detections and non-detections of \ce{H2O2} in a diverse sample of astronomical sources \citep{Bergman:2011,Parise:2014,Liseau:2015}. Modeling studies have shown that the production of peroxide and therefore the gas-phase detectability is very sensitive to temperature. The surface destruction of peroxide was taken into account by rescaling the reaction rate according to experimental data \citep{Du:2012,Parise:2014}. The use of more accurate rate constants may further elucidate the \ce{H2O2} abundance in dark clouds.

In the ISM, the reaction is likely to occur in an environment that consists predominantly of water molecules. In a recent article, we have focused on the reaction between \ce{H2} and \ce{OH} on a crystalline water ice \citep{Meisner:2017}. There, it was shown that the influence of the surface was small and rate constants calculated on a surface differed from those in the gas phase only by a factor three roughly. Moreover, the transition state structure was very similar to that in the gas phase. The \ce{H2O2} molecule, on the other hand, can partake in several hydrogen bonds leading to multiple orientations with respect to the surface. {This does not only affect the binding structure, but the hydrogen bonds can affect the electron density of the transition state and result in a different reactivity than in the gas phase.} Therefore the reaction between a hydrogen atom and the hydrogen peroxide molecule is very suitable to test the influence of hydrogen bonds of the water surface with reactants and products. In a previous paper we have shown that adding one, two, or three water molecules to the system has an influence on the activation energy for the reaction, and, therefore, on the calculated rate constants \citep{Lamberts:2016B}. The low-temperature unimolecular rate constants consequently span a range of approximately an order of magnitude. Here, we extend on the previous studies by the use of a full crystalline water surface as well as amorphous bulk ice to test the influence of both amorphous and bulk effects by means of multiscale quantum mechanical/molecular mechanical (QM/MM) model. The purpose of this paper is thus twofold: on the one hand we provide reaction rate constants that include the effect of tunneling and on the other hand we study the effect of water ice on a reaction of a flexible molecule capable of forming hydrogen bonds.

Previous studies have included amorphicity in various ways and calculated reaction barriers and/or reaction rate constants \citep{Woon:2002,Chen:2011, Rimola:2012,Enrique-Romero:2016, Song:2016}. 
To the best of our knowledge this is the first DFT study that incorporates the effect of a highly amorphous bulk ice and discusses the general implications on reaction dynamics. The importance of the influence of bulk ice on chemical reactivity can be understood by realizing which types of astrochemical models are commonly used. Ideally, a model takes microscopic details into account by keeping track of the location of the molecules in an ice layer, which is the case in kinetic Monte Carlo (kMC) simulations \citep{Cazaux:2010, Lamberts:2014}. Bulk chemistry naturally follows from the location of the reactants, but such models are computationally very expensive and it is not feasible to use them for large reaction networks ($>30$~reactions). Therefore, most models make use of a mean-field rate-equation approach, for instance by defining three phases that are treated separately, but interact with each other, namely gas, surface, and bulk \citep{Hasegawa:1993}. Traditionally, the bulk is rendered chemically inert, since diffusion is limited and therefore chemistry was thought to be inhibited. In more recent astrochemical models, however, bulk chemistry is included explicitly as it can be activated when radicals are created in the ice via, \emph{e.g.}, photochemistry or when diffusion is enhanced through higher temperatures \citep{Garrod:2013, Taquet:2016, Vasyunin:2017}. Bulk chemistry has been experimentally shown to take place for various systems \citep{Noble:2013,Theule:2013,Lamberts:2015}. Since diffusion can indeed be limited this requires the reactants to be present in relatively high concentrations to allow spectral identification of the products.

The paper has been designed as follows, Section~\ref{CompDet} describes the QM/MM setup, the construction of the surface and bulk water ice model, as well as the calculation of reaction rate constants. Section~\ref{RD} is split into two parts, first describing the binding sites, energies, and rate constants on a surface and then in the bulk. We conclude with astrochemical implications in Section~\ref{AstroImpl} and a summary in Section~\ref{Summ}.

\section{Computational details}\label{CompDet}

\subsection{QM/MM setup}

For the description of a reaction on a surface, the computational cost can be greatly reduced by making use of a multiscale modeling setup, namely electrostatic embedded QM/MM {\citep{Warshel:1972,Senn:2007}}. The reactants and the molecules in the direct vicinity of the reactants, determined by the hydrogen bonding structure, are described at the quantum mechanical level (QM), whereas the rest of the ice is described at the force-field level (MM). The QM/MM implementation in Chemshell has been utilized throughout this project {\citep{Sherwood:2003, Metz:2014}}. Visualisation of the system was done using Visual Molecules Dynamics (VMD) \citep{VMD}.

The system to be described at the quantum mechanical level consists of the H atom, the \ce{H2O2} molecule, and several water molecules. Such an amount of atoms ($\gtrsim$ 20) and, more importantly, corresponding gradients and hessians, is too computationally expensive for highly correlated electronic structure theories. Therefore density functional theory (DFT) is used here. A suitable functional and basis set should be able to describe the interaction between hydrogen and hydrogen peroxide, as well as with and between water molecules. We performed a benchmark study for the activation and reaction energies in \citet{Lamberts:2016B} of various functional and basis set combinations against accurate multi-reference wavefunction methods. The best match was obtained for the MPW1B95 functional \citep{Zhao:2004} combined with the MG3S basis set \citep{Lynch:2003}. 

The force field description of the ice water molecules is established via the rigid TIP3P force field {\citep{TIP3P}}. {The limited accuracy of the TIP3P potential is sufficient in our case, since the QM regions are chosen such that they cover the main interaction between the ice molecules and the reactants. In this way we can benefit from its computational efficiency. The MM molecules then serve mainly as boundaries to keep the water ice stable, which is particularly important for bulk ice.} The MM molecules polarize electron density via point charges located on the H and O atoms of the water molecules with a charge of +0.417 and -0.834 e, respectively, taken into account by the Hamiltonian. Simultaneously, the QM molecules influence the MM molecules via electrostatics and their contribution to the bonding and Van der Waals interactions, thus, we follow an electrostatic embedding scheme.

The surface model used is based on a Fletcher surface {\citep{Fletcher:1992}} with ordered protons. Note that the addition of the reactants to the surface, as well as the use of a large initial QM region, locally distort the structure. 
A number of 1151 water molecules constitute a water ice semi-sphere. The stationary points -- binding sites for \ce{H2O2} and transition state structures -- are determined in two stages. In all calculations, the reactants are treated quantum mechanically and are thus always flexible (active). A total of 261 water molecules in a radius of 15~{\AA} from the center of the semi-sphere is set to be flexible. These are largely described on the MM level, but include 20 QM top and second layer water molecules surrounding the reactants. Secondly, the resulting structure is used as input for another geometry optimization where all molecules but a small QM region are frozen. The size of this smaller QM region depends on the number and orientation of hydrogen bonds between the reactants and the surface water molecules and is either 4, 5 or 7, see Table~\ref{SurfEnergies}. Intrinsic reaction coordinates (IRCs) and rate constants are calculated using this smaller active QM region.

The bulk ice has been constructed by a molecular dynamics simulation of a cubic box with $L_\text{box}=31$~{\AA}  with periodic boundary conditions filled with water molecules at room temperature  with the use of NAMD {version 2.8 \citep{NAMD}}. The original box size of 30~{\AA}  has been increased to allow for the formation of rounded edges, which have been turned into a cavity in the center via a coordinate translation of half the box size. In this way, the 826 water molecules constitute an amorphous solid water ice with a cavity in the centre. The total density is thus 0.83 g/cm$^3$, which corresponds to a low-density amorphous ice structure and includes the (empty) cavity, the size of which cannot be unambiguously determined. \ce{H2O2} was placed in the cavity at various starting position and stationary points have been optimized using different QM and active regions for each binding site. A number of 45--54 water molecules around the \ce{H2O2} molecule are set to be flexible including 17 or 18 QM water molecules directly surrounding the reactants, see Table~\ref{BulkEnergies}. Intrinsic reaction coordinates have been calculated using the full QM region, but selecting a set of the 3 closest water molecules that interact with the reactants to be active along with the H atom and the \ce{H2O2} molecule.\\

\begin{table*}[t]
 \centering
 \caption{Binding energies for the adsorption of \ce{H2O2} on the crystalline water surface for the large and small QM regions. The number of hydrogen bonds of the \ce{H2O2} with surrounding \ce{H2O} molecules is indicated. Accompanying activation energies and crossover temperatures in K calculated on the small QM region are also given. $V$ indicates the energy without zero-point energy correction, $E$ gives the vibrationally adiabatic energy difference, all in kJ/mol.}\label{SurfEnergies}
 \begin{tabular}{lrrrrrrr}
 \hline
			& {BS1-s} & {BS2-s} & {BS3-s} & {BS4-s} & {BS5-s} & & {BS6-s} \\
\hline
\# H-bonds		& 3	& 2	& 3	& 3	& 2	& & 2/3	\\
\# QM \ce{H2O}  	& 20	& 20	& 20	& 20	& 20	& & 20	\\
$V_\text{bind}$ 	& -66.1	& -29.4	& -57.8	& -65.7 & -30.3	& & -50.4	\\
ZPE correction		& +11.1	& +7.2	& +7.9	& +9.7	& +5.8	& & +8.8	\\
$E_\text{bind}$ 	& -55.0	& -22.2	& -49.9	& -56.0	& -24.5	& & -41.6	\\
\hline                                                                 
\# QM \ce{H2O}  	& 4	& 4	& 7	& 4	& 4	& 5 	& 7	\\
$V_\text{bind}$ 	& -67.8	& -33.9	& -60.6	& -67.7	& -28.9	& -27.4	& -51.6	\\
ZPE correction		& +11.7	& +9.2	& +9.8	& +11.7	& +7.4	& +7.4	& +9.0	\\
$E_\text{bind}$ 	& -56.1	& -24.7	& -50.8	& -56.0	& -21.5	& -20.0	& -42.6	\\
\hline                                                                 
Deviation $E_\text{bind}$ & 1.1 & 2.5	& 0.9	& 0.0	& 3.0	& 4.5 & 1.0	\\
\hline
 \hline
			& {TS1-s} & {TS2-s} & {TS3-s} & {TS4-s} & {TS5-s} & {TS5-s} & {TS6-s} \\
			& 	& 	& 	& 	& A	& B	& \\
\hline
\# QM \ce{H2O}  	& 4	& 4	& 7	& 4	& 4	& 5	& 7	\\
$V_\text{act.}$ 	& +25.0	& +26.9	& +26.0	& +22.8	& +24.1	& +33.1	& +27.0	\\
ZPE correction		& -0.3	& -2.1	& -2.6	& -0.0	& -1.3	& -1.1	& -1.5	\\
$E_\text{act.}$ 	& +24.7	& +24.8	& +23.4	& +22.8	& +22.8	& +32.0	& +25.5	\\
$T_\text{C}$		& 270.8	& 278.1	& 265.1	& 273.0	& 263.6	& 287.1	& 268.5	\\
\hline
\end{tabular}
\end{table*}

\subsection{Reaction rate constants}

Reaction rate constants are calculated taking tunneling into account explicitly with the use of instanton theory \citep{Miller:1975, Callan:1977}. This is based on statistical Feynman path integrals that incorporate quantum tunneling effects, (see \citet{Kaestner:2014,Richardson:2016}). The instanton is a reaction path that optimizes tunneling, or the minimum action path, which is different from the minimum energy path (MEP). Using the canonic formulation of instanton theory, rate constants can be calculated only for temperatures low enough for the path to spread out, \emph{i.e.}, when tunneling dominates the reaction. This is often the case below the so-called crossover temperature, $T_\text{C} = {\hbar \omega_b}/{2\pi k_\text{B}}$, where $\omega_b$ is the absolute value of the imaginary frequency at the transition state, $k_\text{B}$ is the Boltzmann constant, and $\hbar$ Planck's constant.

Instanton theory as implemented in DL-FIND \citep{Kaestner:2009, Rommel:2011} is used here where the Feynman paths of the instantons are discretized to 40 images (down to 96~K) or 78 images (83 and 74~K). Instanton geometries are converged until the gradient is below $10^{-9}$ atomic units. 

{Throughout the full paper we assume that the reaction takes place between thermalized species. The species meet via diffusion and hence the reaction follows the Langmuir-Hinshelwood mechanism. The relevant rate constant to calculate here describes the decay of the encounter complex. Thus, we calculate unimolecular reaction rate constants excluding the probability to meet, leading to a unit of s$^{-1}$.}
Since for all systems considered here rotational motion is restricted, the rotational partition function is assumed to be constant during the reaction, \emph{i.e.}, the rotational partition function of the PRC is assumed to be equal to that of the transition state. Similarly, a rotational symmetry factor is not taken into account either.

\section{Results and Discussion}\label{RD}

\subsection{Surface}

\begin{figure*}[t]
\centering
    TS1-s \hspace{0.3\textwidth} TS5-s A  \\
    \includegraphics[width=0.3\textwidth]{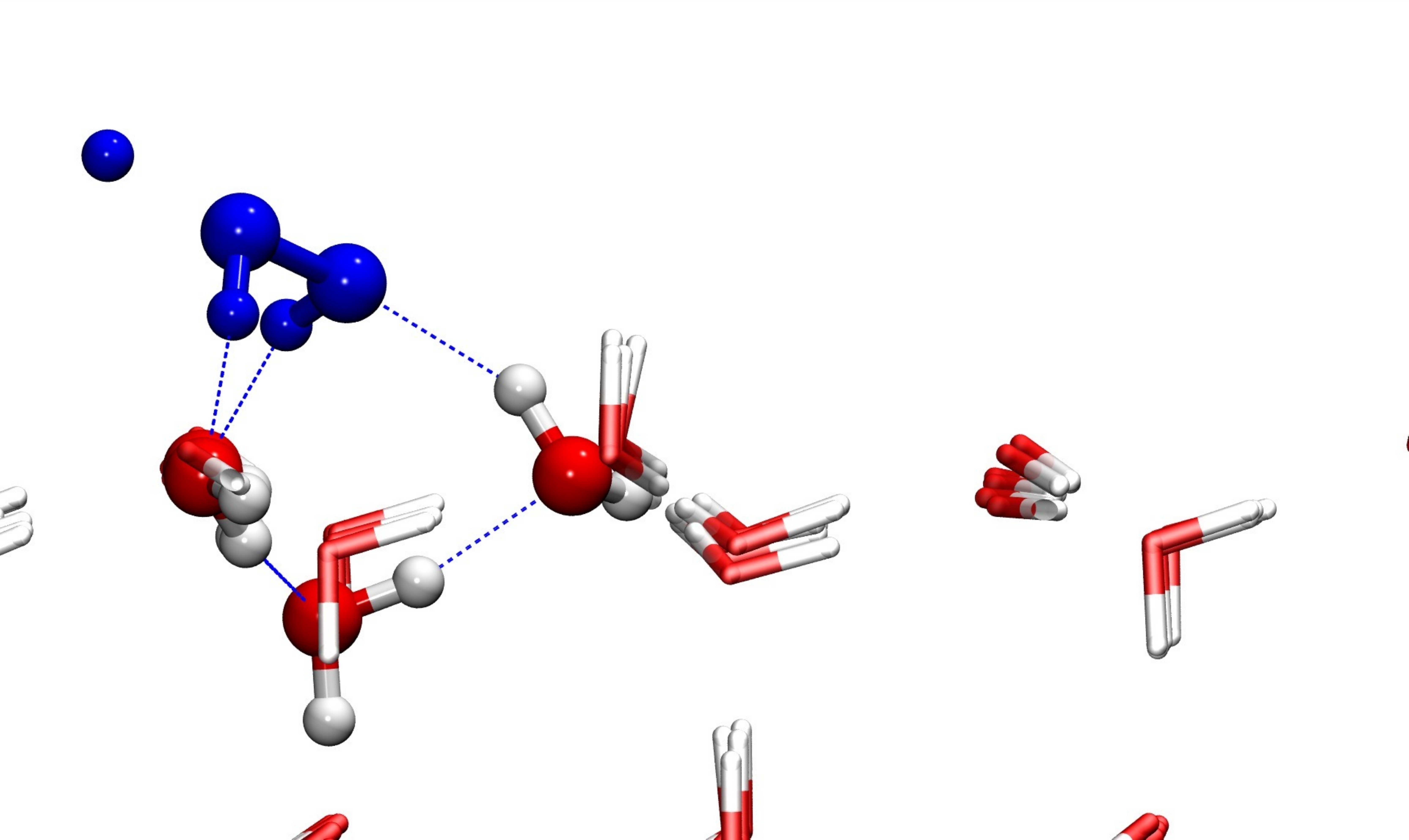}   \hspace{15pt}
    \includegraphics[width=0.3\textwidth]{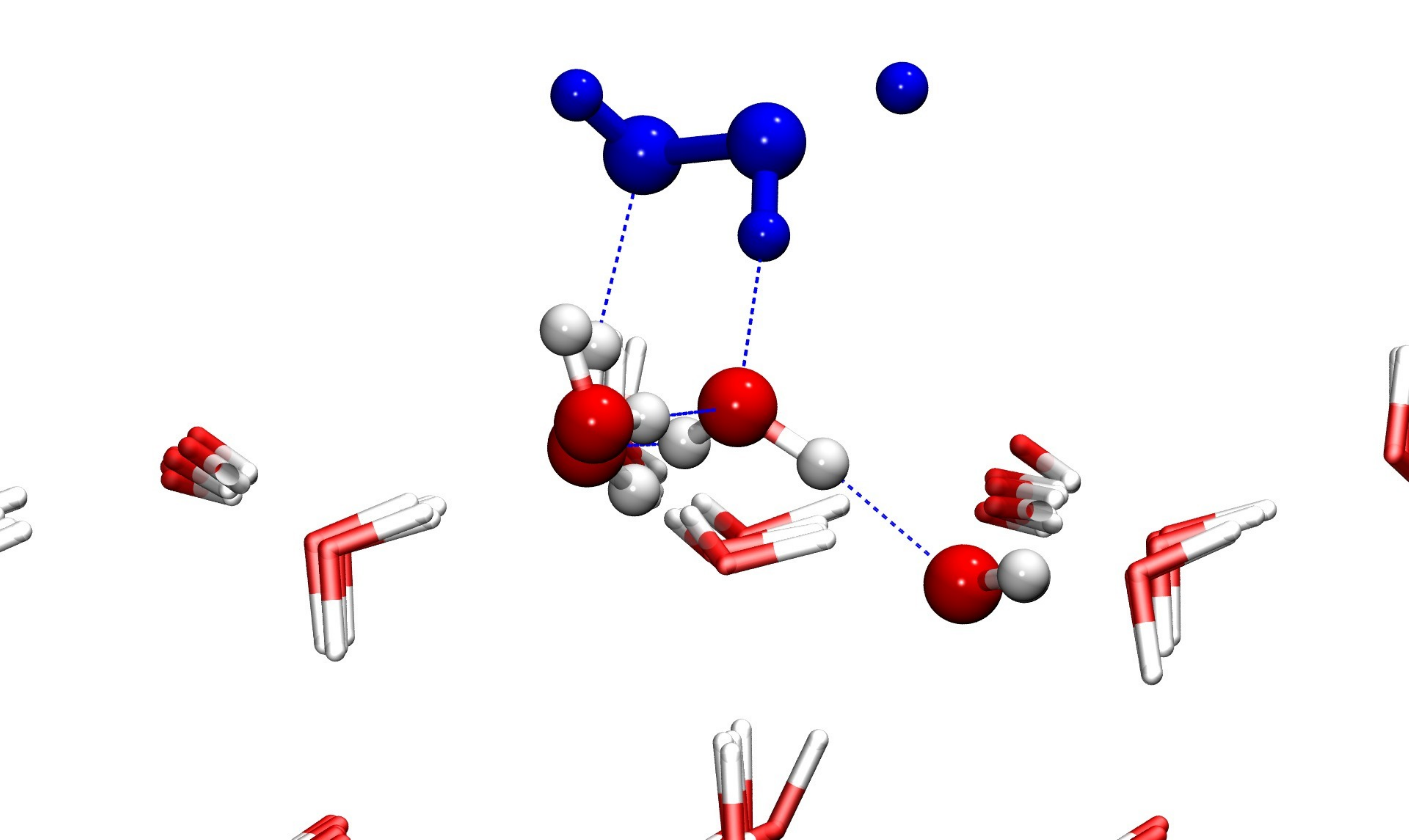}   \vspace{5pt} \\
    \includegraphics[width=0.3\textwidth]{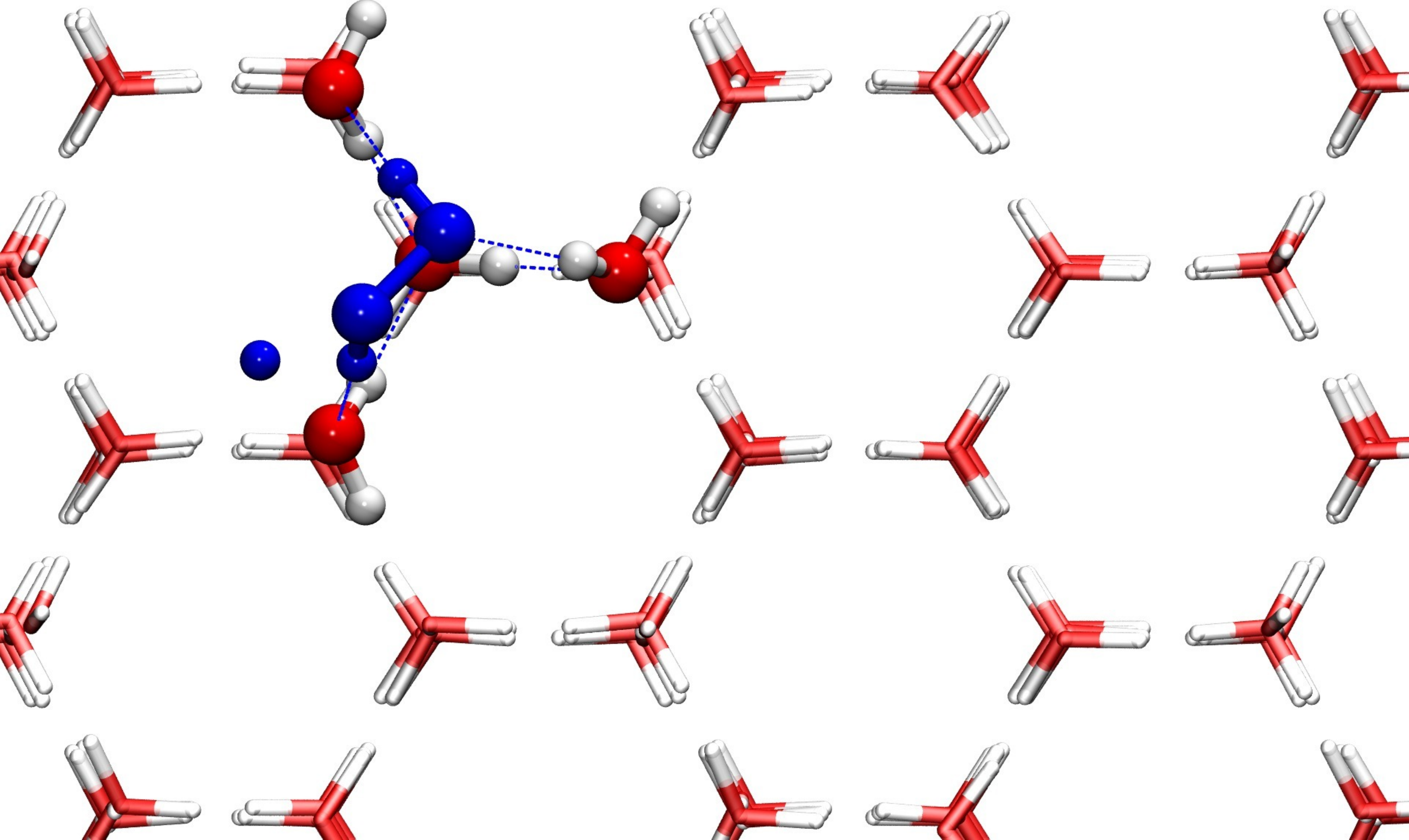}   \hspace{15pt}
    \includegraphics[width=0.3\textwidth]{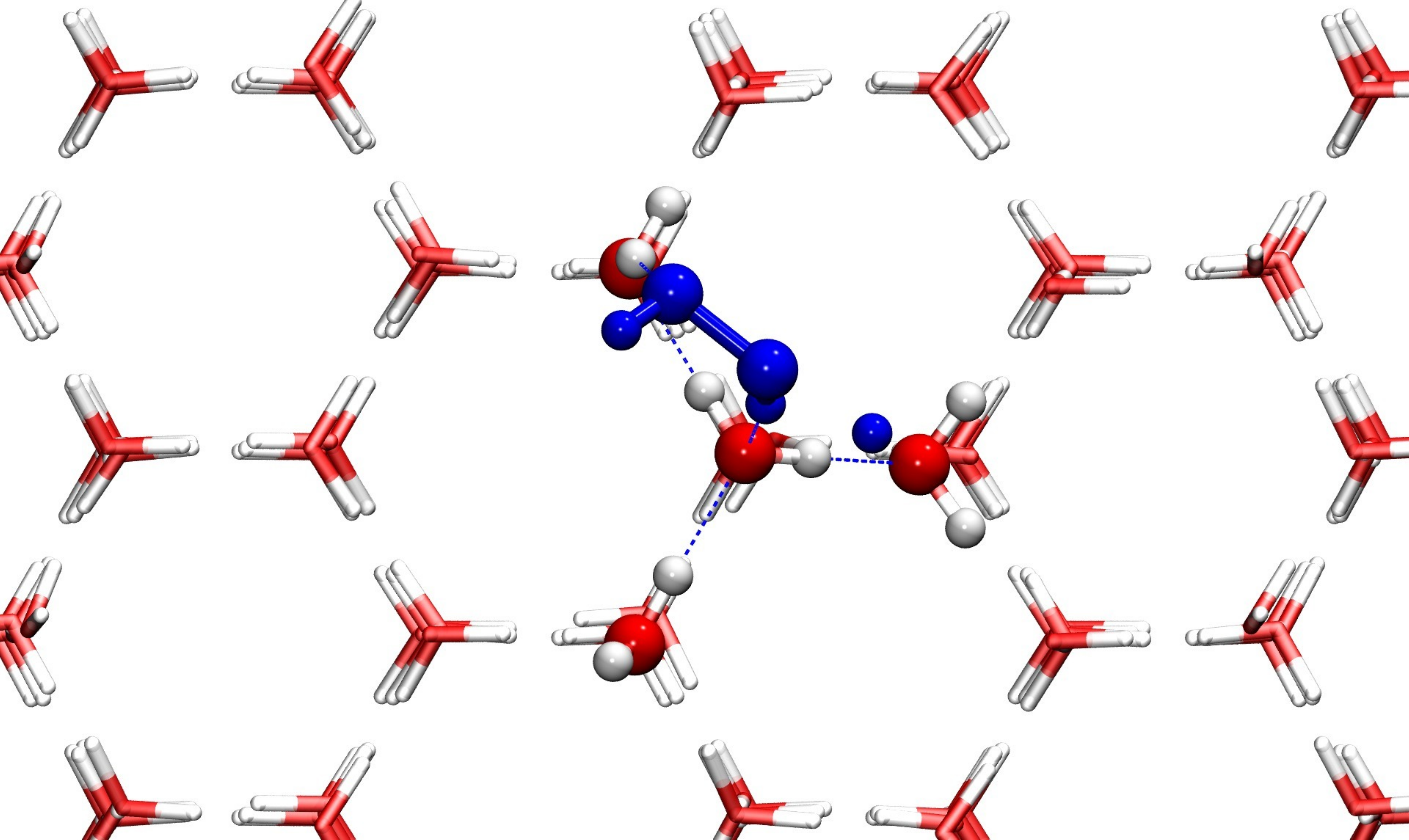}   \vspace{15pt} \\
    TS1-b A \hspace{0.3\textwidth} BS2-b   \\
    \includegraphics[width=0.34\textwidth]{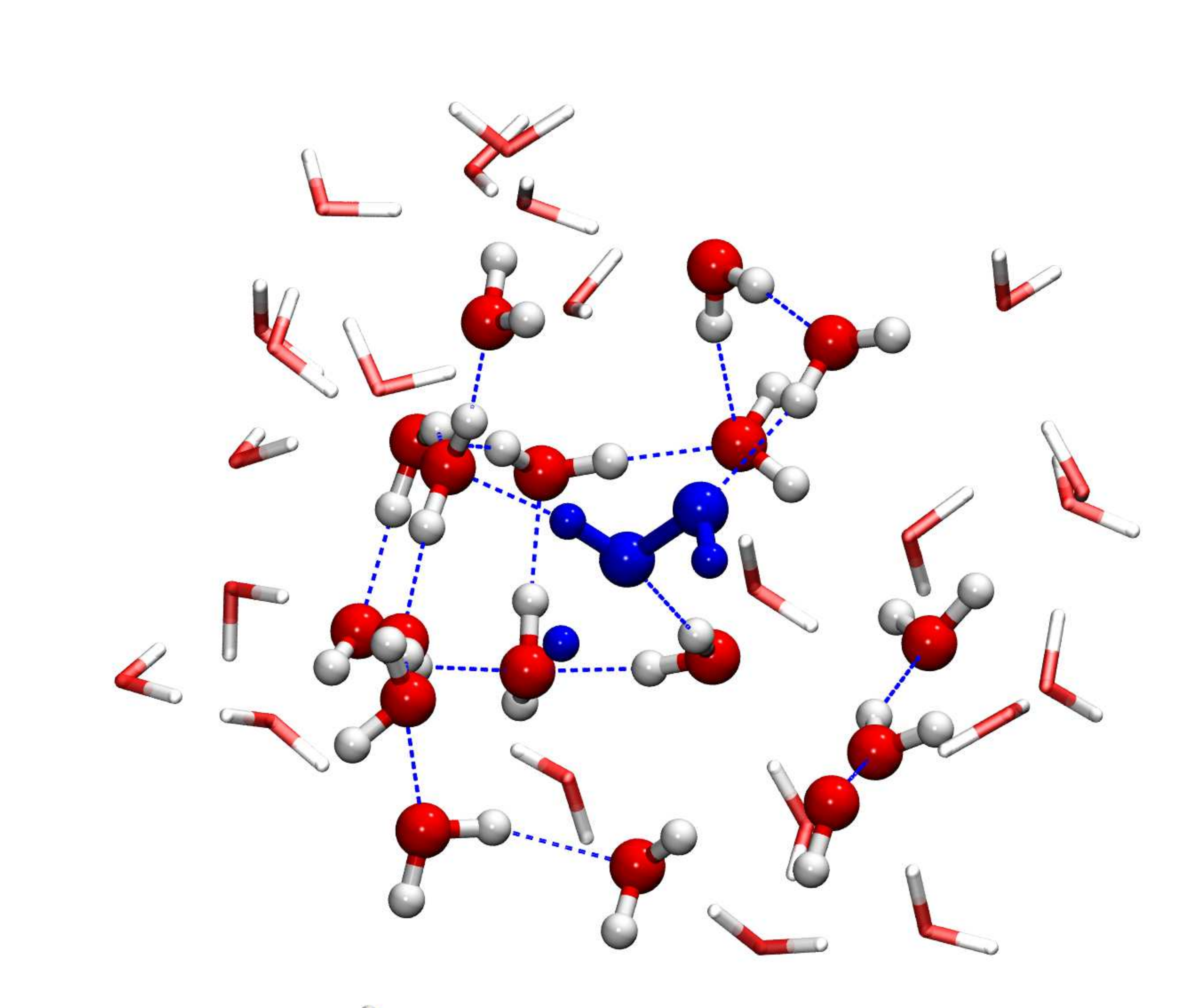}   
    \includegraphics[width=0.34\textwidth]{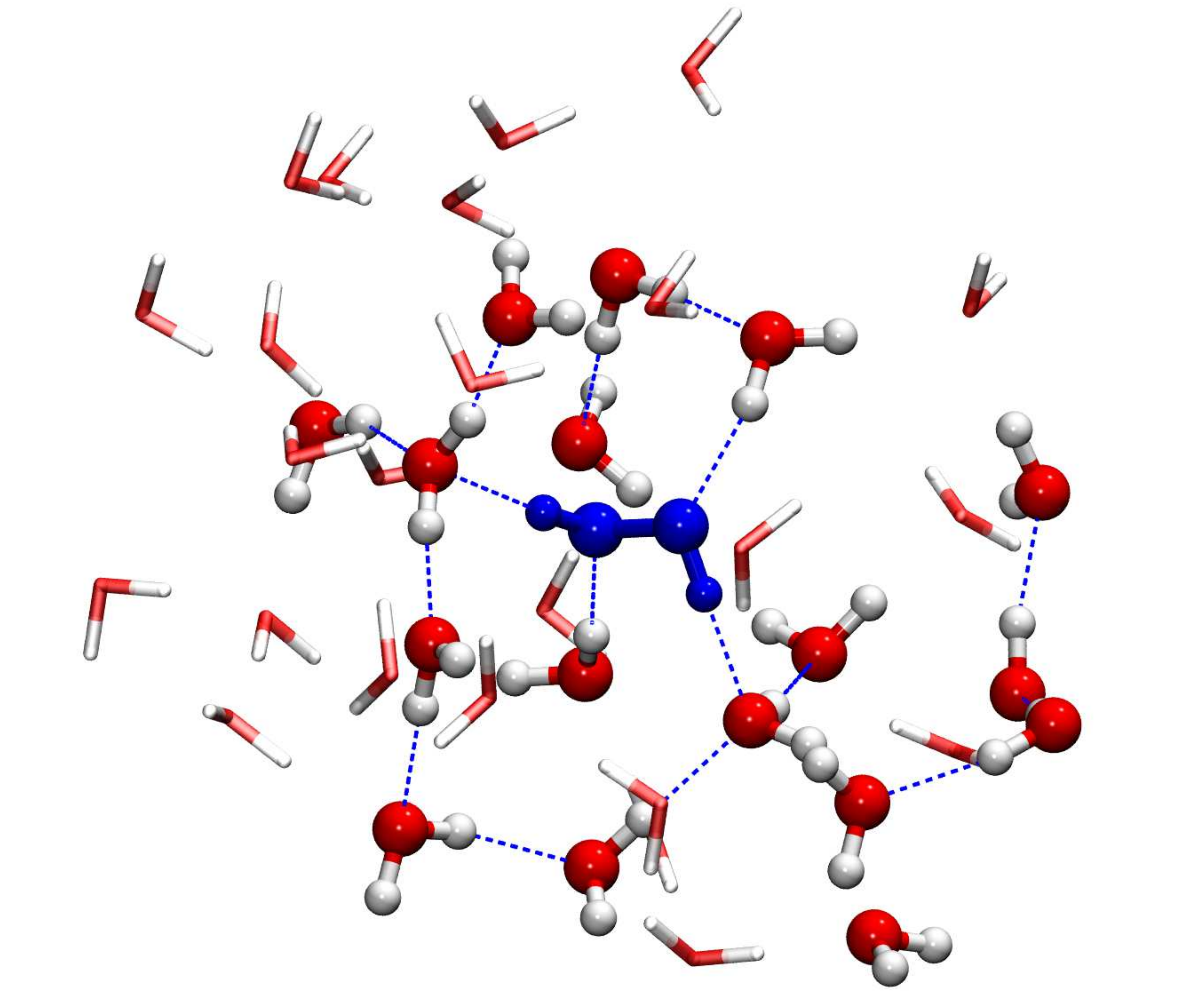}   
    \caption{Transition state structures for TS1-s, TS5-s (surface), and TS1-b A and the binding mode orientation for BS3-b (bulk). Structures on the surface are presented both with a top and side view, those in the bulk have been oriented such to obtain the clearest view of the reactants. The QM molecules are depicted as ball-and-stick, while the MM molecules are given as translucent sticks. The reactants and the hydrogen bonding network within the QM region are given in blue. In the representation shown here, the largest part of the surface has been removed for clarity.}
    \label{TSall}
\end{figure*}

We focus on six different binding sites and accompanying transition states on the crystalline surface, rather than presenting an exhaustive list of different binding modes and energies. Also, although other binding sites have been found, throughout various stages of reoptimization and/or transition state searches, they resulted in structures similar or equal to the six given here. The binding energies and zero-point energy (ZPE) corrections calculated with both large and small QM regions are summarized in Table~\ref{SurfEnergies}. They are defined with respect to seperated \ce{H2O2} and \ce{H2O} surface, \emph{i.e.}, negative values correspond to exothermic binding. The deviation between the large and small QM regions is at most 4.5 kJ/mol for $E_\text{bind}$, generally much lower. Given these relatively small deviations and since the use of a large QM region restricts the possibility of doing instanton rate constant calculations, only the small QM region will be used in the following. 

Even on the crystalline surface the binding energies show a large spread, from -56.0 to -20.0 kJ/mol. Furthermore, the binding energy can be roughly related to the number and type of hydrogen bonds formed between the hydrogen peroxide molecule and the water molecules of the surface, see also Fig.~\ref{TSall}. In BS2-s and BS5-s \ce{H2O2} has one accepting and one donating H-bond with surface water molecules, while in the other structures it has three H-bonds. More specifically, BS1-s and BS4-s, have two accepting and one donating hydrogen bond, but BS3-s and BS6-s have one accepting and two donating hydrogen bonds. In Fig.~\ref{TSall} two transition state structures, TS1-s and TS5-s A, are depicted indicating which water molecules are part of the small QM region. Note that in the case of BS5-s, there are two options for attacking the hydrogen peroxide molecule, resulting in two transition state structures. For the other binding modes, however, either the required proximity of the H atom to surface water molecules or a donating hydrogen bond to the peroxide O atom prevent a reaction from taking place on both sides.

In Table~\ref{SurfEnergies} the values for the activation energy for the reaction \ce{H + H2O2 -> H2O + OH} are given. 
They have been calculated with respect to the PRC on the surface and range between 22.8 and 25.5 kJ/mol with one additional value at 32.0 kJ/mol. {In the PRC structures the hydrogen atom is always close to the \ce{H2O2} molecule rather than directly on the surface, similar to the gas-phase PRC found previously \citep{Lamberts:2016B}. } These activation energies cannot be related to the binding energy of the original \ce{H2O2} site. This can be compared to the values of `$E_\text{a}$ unimol' in Table~2 of \citet{Lamberts:2016B} of gas-phase and cluster models, with values ranging between 24.2 and 26.6 with an additional value of 30.4 kJ/mol. In Fig.~\ref{AllRC} unimolecular rate constants calculated with instanton theory down to 83~K or 74~K are depicted for TS1-s, TS3-s, TS4-s, TS5A-s, and TS5B-s along with rate constants calculated in \citet{Lamberts:2016B}. A somewhat larger spread can be observed, \emph{i.e.}, {approximately 2 orders of magnitude, although most rate constants are somewhat higher than in the cluster model, in accordance with the calculated activation energies.} Furthermore, the barrier for TS4-s is small, but the calculated intrinsic reaction coordinate (IRC) for this binding site shows that it is relatively broad, and therefore the rate constant decreases more steeply with lower temperature than for the other cases. The barrier for TS2-s on the other hand is the narrowest leading to a higher rate constant at lower temperatures. The curves given are somewhat noisy resulting from the small number of images used. The computational cost of instanton rate calculations, where hessians need to be calculated for each image, prevents the use of a highly resolved instanton path. 

\subsection{Bulk}

\begin{table*}[t]
 \centering                                                                                                                   
 \caption{Binding energies for the adsorption of \ce{H2O2} inside an amorphous water surface calculated with a large QM region. The number of hydrogen bonds of the \ce{H2O2} with surrounding \ce{H2O} molecules is indicated. Accompanying activation energies and crossover temperatures in K are also given for the cases where a transition state was found. $V$ indicates the energy without zero-point energy correction, $E$ gives the vibrationally adiabatic energy difference, all in kJ/mol.}\label{BulkEnergies}
 \begin{tabular}{lrrrrrrr}
 \hline
			& {BS1-b} &  	& {BS2-b} & {BS3-b} & {BS4-b} \\
\hline
\# H-bonds		& 3	& 	& 4	& 3	& 1	\\
\# QM \ce{H2O}  	& 17	& 	& 17	& 17	& 18	 \\
\# active \ce{H2O}	& 53	& 	& 45	& 54	& 52	\\
$V_\text{bind}$ 	& -40.2	& 	& -55.9	& -52.2	& -33.3	\\
ZPE correction		& +12.6	& 	& +12.4	& +10.4	& +14.5	 \\
$E_\text{bind}$ 	& -27.6	& 	& -43.5	& -41.8	& -18.7	\\
\hline                                                           
\hline			
			& {TS1-b} & {TS1-b} & 	& {TS3-b} & {TS4-b} \\
			& A	& B	& 	& 	& 	\\
\hline
\# H-bonds		& 3	& 4	& 	& 3	& 1	\\
$V_\text{act.}$ 	& +27.6	& +32.2	& 	& +24.3	& +25.2	\\
ZPE correction		& -4.0	& -0.5	& 	& -1.3	& -4.4	\\
$E_\text{act.}$ 	& +23.6	& +31.7	& 	& +23.0	& +20.8	\\
$T_\text{C}$		& 272.8	& 284.5	& 	& 265.2	& 264.4	\\
\hline
\end{tabular}
\end{table*}

Again, we investigated several binding sites also in bulk ice in order to understand the differences and similarities between the same reaction that takes place on an ice surface or inside an ice.

For the amorphous bulk ice, binding sites and transition states have been identified inside the cavity. In this case more QM water molecules are required to obtain a good description of the system and the size of the QM region can thus not be reduced. The binding and activation energies are summarized in Table~\ref{BulkEnergies} and one transition state structure, TS1-b, and one \ce{H2O2} binding mode, BS2-b, are depicted in Fig.~\ref{TSall}. 

The structure of the ice cavity without an adsorbate is based on various hydrogen bonding interactions that are disturbed or altered by the addition of a molecule. The binding energy given in the Table is therefore a combination of ice restructuring and new interactions established with hydrogen peroxide. The \ce{H2O2} binding site BS1-b has two accepting and one donating hydrogen bond, for BS3-b it is the other way around, and BS4-b has (only) one donating hydrogen bond. Note that for BS2-b in particular, the orientation of the four hydrogen bonds (two accepting and two donating) prevent a transition state from being formed, see Fig.~\ref{TSall}. 
Although the binding energies on the surface and in the bulk cannot be directly compared to each other, the number of hydrogen bonds does indicate that a different type of binding of the \ce{H2O2} molecule takes place. On the flat crystalline surface, the maximum number of hydrogen bonds is restricted to three, whereas in a more amorphous surrounding, be it surface or bulk, more interactions can occur. This may result in structures that are so stable that attack by H is impeded, for instance in the case of BS2-b. Similar effects have been previously found for the reaction system \ce{CH3 + HCO} by \citet{Enrique-Romero:2016}.

\begin{figure}[t]
 \centering
 \includegraphics[width=0.46\textwidth]{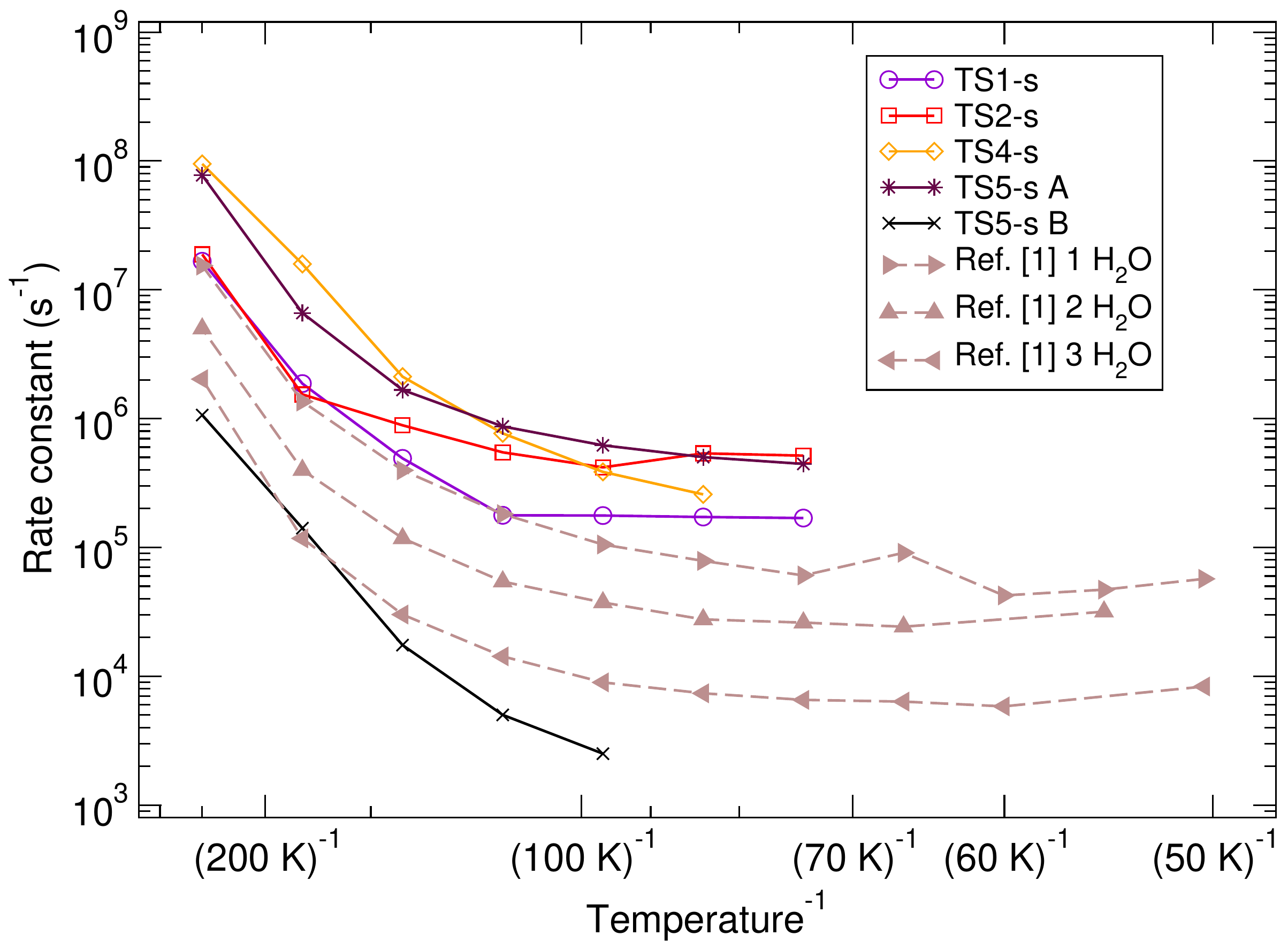}
 \caption{Unimolecular rate constants for five binding sites on the crystalline surface along with previously calculated values for the gas-phase reaction with spectator \ce{H2O} molecules, [1] \citet{Lamberts:2016B}.}\label{AllRC}
\end{figure}

\begin{figure}[t]
 \centering
 \includegraphics[width=0.46\textwidth]{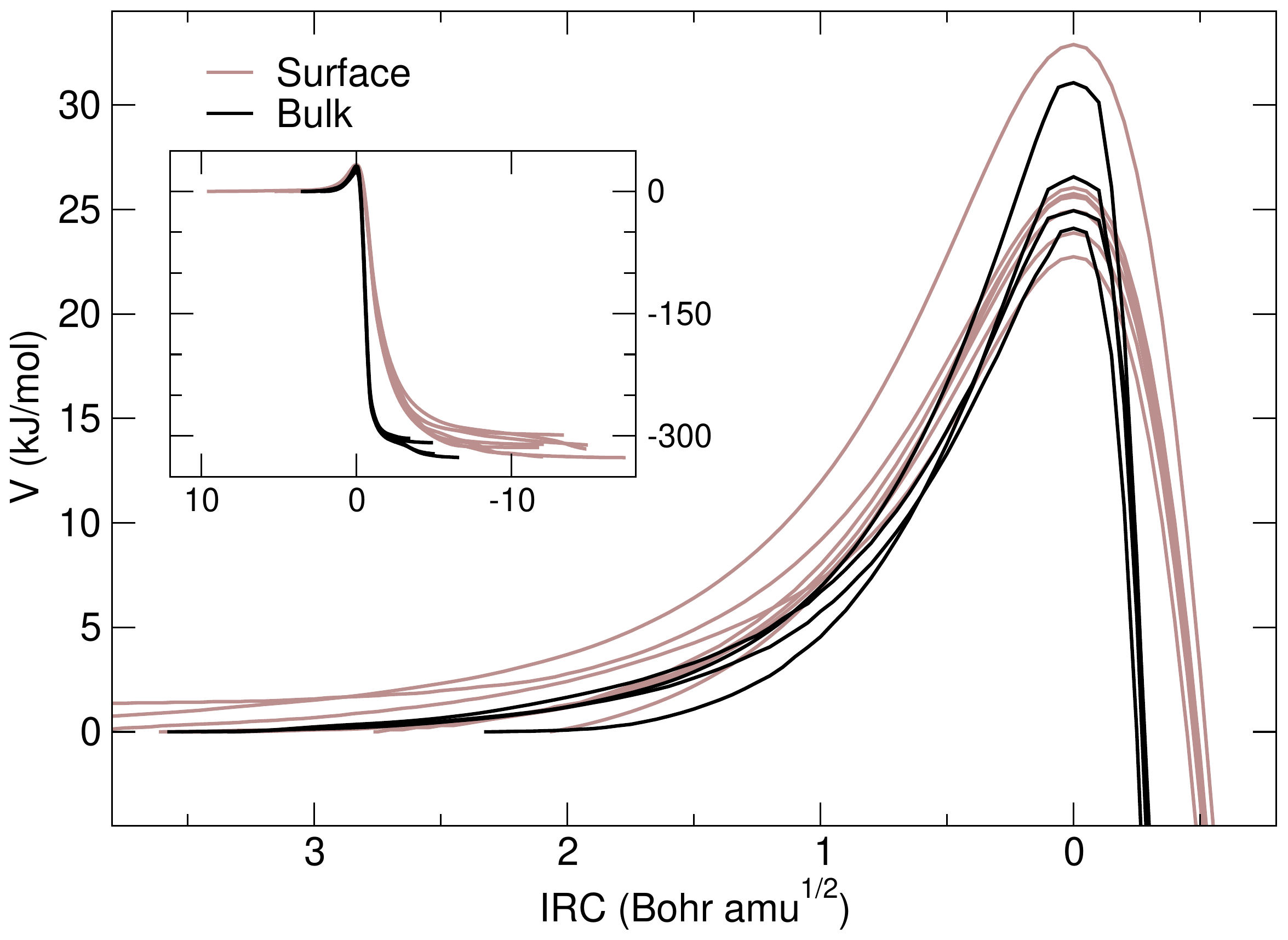}
 \caption{Intrinsic reaction coordinates calculated for all transition states found on the surface (brown curves) and in the bulk ice (black curves). The inset provides the full IRCs from which also the reaction energy can be determined.  }\label{IRCSurf}
\end{figure}

The activation energies span a range of 20.8--31.7 kJ/mol. Again, as was the case for the small water clusters and for the surface studies, there is only one transition state with an activation energy $> 30$~kJ/mol. In general in this study, the lower barriers correspond to early transition states, whereas the higher barriers correspond to late transition states. This is consistent between the previous cluster study and the surface and bulk results from the present study. It is reasonable to assume that the higher the barrier, the less likely it is for a reaction to occur as a result of the competition with diffusion of the hydrogen atom on the surface. Without considering these `outlier' cases, comparing the activation energies results in the following ranges:
\begin{center}
\begin{tabular}{|lllll|}
\hline
& \multicolumn{1}{c}{Clusters} &  \multicolumn{1}{c}{Surface} &  \multicolumn{1}{c}{Bulk} & \\
$E_\text{act.}$ & 24.2--26.6 & 22.8--25.5  & 20.8--23.6 & kJ/mol. \\
\hline
 \end{tabular}
\end{center}
Although the ranges overlap, they are clearly shifted and this impacts on the rate constant as can also be seen in Fig.~\ref{AllRC}. It is to be expected that rate constants in the amorphous bulk would thus be even slightly higher than those on the surface. Rate constants at low temperature are not only determined by the height of the barrier, however, but also by the barrier width. Unfortunately, the required size of the QM region prevents the use of instanton theory for calculating reaction rate constants. Therefore, IRCs have been calculated to obtain additional information about the barrier width. The IRCs for all transition state structures are depicted in Fig.~\ref{IRCSurf}. The height of the barrier corresponds directly to the values without ZPE correction, \emph{i.e.}, $V$, mentioned in Tables~\ref{SurfEnergies} and~\ref{BulkEnergies}. All bulk IRCs are shorter and narrower than those on the surface both towards the reactant and product sides. This is the result of the restricted freedom for molecular reorientation in the highly amorphous surrounding. This in turn is expected to lead to an additional enhancement of the rate constants in the bulk ice at low temperature with respect to those on the crystalline surface.

\section{Astrochemical Implications}\label{AstroImpl}

The reaction \ce{H + H2O2 -> H2O + OH} is especially fit for studying the possible influence of hydrogen bonds and strong adsorbate-surface interaction on chemical reactivity on icy grains. This study can thus be seen as broader than the calculation of rate constants on a surface, as it emphasizes on the possible effects of an amorphous and bulk environment. With this in mind, below we highlight several astrochemical implications.

It is important to take into account that in the ISM a \ce{H2O2} molecule would not adsorb on the surface from the gas phase, but would rather be formed directly in the water surrounding. Therefore, the most likely binding mode cannot be trivially determined by the strongest (lowest) binding energy. In fact, it should be determined by a combination of (1) the binding energy, (2) the competition with diffusion, and (3) the dissipation of energy (after formation) of the precursor atoms and molecules, \emph{i.e.}, O, \ce{O2}, OH, and \ce{HO2}. These precursors form during the same dark cloud stage as when the water ice is formed, and the total process is thus an intricate interplay of intermolecular interactions.

The range of binding energies obtained here for the amorphous surrounding (bulk) is in good agreement with values commonly used in models of about $\sim -45$~kJ/mol \citep{Taquet:2013,Garrod:2008b} or $\sim -10$~kJ/mol per neighbor or next-nearest neighbor \citep{Lamberts:2013, Garrod:2013}. However, usually in rate-equation models a single value for the binding energy is used, while we find a rather broad distribution of binding energies, as has been the case in previous studies for various species on a water surface \citep{Karssemeijer:2014,Karssemeijer:2014b,Song:2016,Asgeirsson:2017,Senevirathne:2017}. {Experimental values are not well constrained as a result of the decomposition of \ce{H2O2} molecules on the metal surfaces of the deposition lines and the experimental main chamber, thus complicating both the deposition of a pure \ce{H2O2} ice and the quantification during temperature programmed desorption experiments.}

There is no correlation between the binding energy (or the number of hydrogen bonds) and the height of the activation energy. Furthermore, knowing that the surface and bulk reduce the barrier height, the previously suggested modified Arrhenius expression \citep{Lamberts:2016B} should be considered as a \underline{lower limit} to the rate constant on and in interstellar ices 
$$ k_{\text{LH}} > 1.51 \times 10^{10} \left(\frac{T}{300}\right)^{0.86}\exp\left(-1750\frac{T+180}{T^2+180^2}\right) \;\text{s}^{-1}. $$
As a result of the classical catalytic effects of the ice surface and bulk on the rate constant, the value of $k_\text{LH}$ can be up to two orders of magnitude higher. 
Comparing the rate constants obtained here to those used in literature shows that the value used by \citet{Du:2012} has been much too high, by about 4 orders of magnitude. On the other hand, the probability used by \citet{Taquet:2013,Taquet:2016} multiplied with a trial frequency of $10^{12}$ s$^{-1}$ yields rate constants that are on the low side of the range obtained here. Since single reaction pathways are always linked to many other processes within the model used, it is not a priori obvious to say how our new values will change the outcome of their models. 

Concerning the kinetic isotope effects (KIEs), the decrease in the activation energy of the reaction \ce{D + H2O2 -> HDO + OH} with respect to \ce{H + H2O2 -> HDO + OH} is 1.03 kJ/mol and 0.85 kJ/mol for B1-s and B3-s, respectively. {Those energy differences stem exclusively from the zero point energy, \emph{i.e.}, the difference in mass in the frequency calculations. Thus, they can be considered reliable even though they are smaller than the typical accuracy of DFT expected for reaction energies or barriers.} Also for these two cases the rate constants on the surface are enhanced by about one and a half orders of magnitude, leading us to conclude that although the rate constants increase, the effect on the KIE should be minimal with respect to the previously published values. 

Finally, we want to stress once more that regardless of the influence of intermolecular interaction between the surface molecules and the reactants, an interstellar icy grain effectively leads to an increased concentration of reactants with respect to that in the gas phase and allows for heat dissipation of the exothermicity of the reaction. In this way it acts as a catalyst for a multitude of reactions that could not take place in the gas phase. At the same time, the interaction with the surface may also prevent chemical reactions from taking place, depending on the orientation of the reactants on the surface or in the bulk ice.

\section{Summary}\label{Summ}

The main findings for our study of the effect of a crystalline water surface and amorphous bulk ice on the reaction \ce{H + H2O2 -> H2O + OH} are listed below. This project may be seen as testing various aspects of the existence of the influence of water ice  or lack thereof on reactivity. As a result of the larger flexibility of the \ce{H2O2} molecule we found a clear influence of the ice surrounding on the reactivity. Moreover, the enhancement of the rate constants on the surface and in the bulk has been rationalized within a hydrogen bonding framework and should be kept in mind when considering large molecules on ice surfaces.

\begin{itemize}
\item Binding energies to the surface cannot be captured in a singly value, but do vary with the number of hydrogen bonds with neighboring molecules.
\item Amorphous surrounding can lead to more hydrogen bonding and hence prevent reaction partners to meet.
\item Binding energies are not correlated to the activation energies of the reaction taking place on the same binding site.
\item Rate constants related to the Langmuir-Hinshelwood mechanism including the effect of tunneling on the surface are somewhat higher than those calculated with the same method for the cluster model.
\item Rate constants in highly amorphous and/or bulk environments are expected to be again higher than those calculated on a crystalline surface.
\item Thus, the previous parametrization of the modified Arrhenius expression \citep{Lamberts:2016B} is better seen as a lower limit for the rate constant.
\end{itemize}

\acknowledgments
The authors acknowledge support for computer time by the state of Baden-W\"{u}rttemberg through bwHPC and the Germany Research Foundation (DFG) through grant no. INST 40/467-1FUGG. 
This project was financially supported by the European Union's Horizon 2020 research and innovation programme (grant agreement No. 646717, TUNNELCHEM). TL wishes to acknowledge the Alexander von Humboldt Foundation for generous support.  We thank Max Markmeyer and Jan Meisner for useful discussions on the IRC paths.

\bibliography{biblioH2O2surf}

\end{document}